\documentclass[preprint,superscriptaddress,floatfix,longbibliography]{revtex4-1}

\usepackage{graphicx}
\usepackage{dcolumn}
\usepackage{bm}
\usepackage{soul}
\usepackage{amssymb}
\usepackage{microtype}
\usepackage{xfrac}
\usepackage{gensymb}
\usepackage{xcolor}
\usepackage{amsmath}
\usepackage{siunitx}

\begin{document}

\title{Quantum critical point in the itinerant ferromagnet Ni$_{1-x}$Rh$_x$}

\author{C.-L. Huang} 
\email[]{clhuang1980@gmail.com}
\affiliation{Department of Physics and Astronomy, Rice University, Houston, Texas 77005, USA}

\author{A.~M.~Hallas}
\affiliation{Department of Physics and Astronomy, Rice University, Houston, Texas 77005, USA}
\affiliation{Department of Physics and Astronomy and Quantum Matter Institute, University of British Columbia, Vancouver, British Columbia, Canada V6T 1Z1}

\author{K.~Grube}
\affiliation{Institut f\"{u}r Quantenmaterialien und -technologien, 76021 Karlsruhe, Germany}

\author{S.~Kuntz}
\affiliation{Institut f\"{u}r Quantenmaterialien und -technologien, 76021 Karlsruhe, Germany}

\author{B.~Spie\ss}
\affiliation{Department of Physics and Astronomy, Rice University, Houston, Texas 77005, USA}
\affiliation{Department of Chemistry, Johannes Gutenberg-University Mainz, 55131 Mainz, Germany}

\author{K.~Bayliff}
\affiliation{Department of Chemistry, Rice University, Houston, Texas 77005, USA}

\author{T. Besara}
\affiliation{National High Magnetic Field Laboratory, Tallahassee, Florida 32310, USA}
\affiliation{Department of Physics, Astronomy, and Materials Science, Missouri State University, Springfield, Missouri 65897, USA}

\author{T. Siegrist}
\affiliation{National High Magnetic Field Laboratory, Tallahassee, Florida 32310, USA}

\author{Y.~Cai}
\affiliation{Department of Physics and Astronomy, McMaster University, Hamilton, Ontario L8S 4M1, Canada}

\author{J.~Beare}
\affiliation{Department of Physics and Astronomy, McMaster University, Hamilton, Ontario L8S 4M1, Canada}

\author{G.~M.~Luke}
\affiliation{Department of Physics and Astronomy, McMaster University, Hamilton, Ontario L8S 4M1, Canada}
\affiliation{TRIUMF, 4004 Wesbrook Mall, Vancouver, B.C. V6T 2A3, Canada}

\author{E.~Morosan}
\affiliation{Department of Physics and Astronomy, Rice University, Houston, Texas 77005, USA}

\date{\today}

\begin{abstract}
We report a chemical substitution-induced ferromagnetic quantum critical point in polycrystalline Ni$_{1-x}$Rh$_x$ alloys. Through magnetization and muon spin relaxation measurements, we show that the ferromagnetic ordering temperature is suppressed continuously to zero at $x_{crit} = 0.375$ while the magnetic volume fraction remains 100\% up to $x_{crit}$, pointing to a second order transition. Non-Fermi liquid behavior is observed close to $x_{crit}$, where the electronic specific heat $C_{el}/T$ diverges logarithmically, while immediately above $x_{crit}$ the volume thermal expansion coefficient $\alpha_{V}/T$ and the Gr{\"u}neisen ratio $\Gamma = \alpha_{V}/C_{el}$ both diverge logarithmically in the low temperature limit, further indication of a ferromagnetic quantum critical point in Ni$_{1-x}$Rh$_x$. 
\end{abstract}

\maketitle

A quantum critical point (QCP) occurs when a phase transition is continuously suppressed to zero temperature. The intense quantum fluctuations in the vicinity of a QCP profoundly alter a material's electronic properties, resulting in non-Fermi liquid behavior and, in some cases, unconventional superconductivity \cite{Loehneysen2007,schuberth2016emergence}. The most ubiquitous QCP separates an antiferromagnetically ordered state from one in which quantum fluctuations disrupt the order. Notable examples are found among heavy fermion systems \cite{Stewart2001,Stewart2006,Loehneysen2007}. QCPs in ferromagnetic (FM) metals have proven far more elusive \cite{Manuel2016}. It is now understood that a FM QCP is inherently unstable and can survive only in rare circumstances~\cite{Belitz1999PRL}. In this work, we report the discovery of a FM QCP in Ni$_{1-x}$Rh$_x$, as evidenced by  (i) a second-order phase transition up to the critical concentration $x_{crit}$, and (ii) divergence of the electronic specific heat coefficient $C_{el}/T$, the volume thermal expansion $\alpha_{V}/T$, and the Gr{\"u}neisen ratio $\Gamma = \alpha_{V}/C_{el}$. The dilution of the $d-$electron magnetic sublattice as the tuning parameter to induce a FM QCP opens a new route for exploring FM quantum criticality and possible new collective phases near the QCP, such as unconventional superconductivity \cite{Ran2019}.

FM QCPs are revealed via chemical substitution in Zr$_{1-x}$Nb$_{x}$Zn$_2$ \cite{Sokolov2006}, SrCo$_2$(Ge$_{1-x}$P$_x$)$_2$ \cite{Jia2011}, YbNi$_4$(P$_{1-x}$As$_x$)$_2$ \cite{Steppke2013}, and (Sc$_{1-x}$Lu$_x$)$_{3.1}$In \cite{Svanidze2015}. The disorder effect is minimal or negligible in these systems. For SrCo$_2$(Ge$_{1-x}$P$_x$)$_2$, the QCP is induced by the breaking of dimers \cite{Jia2011}. However, the exact mechanism responsible for the FM QCP in the other three systems remains unclear. In most other FM metals, the QCP is preempted when the continuous (second-order) transition as a function of non-thermal control parameter either becomes discontinuous (first-order), or the ferromagnetism is replaced by a spatially-modulated ordered state \cite{Butch2009,abdul2015modulated,Manuel2016,Taufour2016,Kaluarachchi2017}. Theoretical work by Belitz, Kirkpatrick, and Vojta (BKV) has proposed a route towards a FM QCP by long-range effective spin interactions that occur in the presence of quenched disorder~\cite{Belitz1999PRL,Sang2014,Kirkpatrick2015}. A handful of FM QCPs have been identified as candidates for this phenomenology, including UCo$_{1-x}$Fe$_{x}$Ge~\cite{Huang2016}, (Mn$_{1-x}$Fe$_x$)Si~\cite{Goko2017}, NiCoCr$_x$~\cite{Sales2017}, and Ce(Pd$_{1-x}$Ni$_{x}$)$_2$P$_2$ \cite{Lai2018}, where disorder is inherently introduced by the chemical substitution. In most of these systems, the proposed existence of a QCP is based on either divergence of some thermodynamic parameters \cite{Huang2016,Sales2017,Lai2018} or the second order nature of the transition \cite{Goko2017}. However, the unambiguous identification of a QCP requires that both these criteria be fulfilled. This point is exemplified by disordered Sr$_{1-x}$Ca$_x$RuO$_3$, for which a QCP can be ruled out because the transition at $T = 0$ is first order~\cite{Uemura2006}, and yet, quantum critical scaling is still observed~\cite{Huang2015}. Thus, in order to unambiguously identify a FM QCP it is essential that both thermodynamic signatures of quantum fluctuations \textit{and} second-order behavior be observed simultaneously. 

 Elemental Ni, which has a simple face-centered cubic structure, is known to order ferromagnetically below its Curie temperature $T_{\rm C} = 627$~K \cite{Kraftmakher1997}. Upon alloying with Rh, the $T_{\rm C}$ of Ni$_{1-x}$Rh$_x$ is quickly suppressed \cite{Bucher1967}. Ni$_{1-x}$Rh$_x$ has more configuration entropy than pure Ni \cite{Yeh2004}. Also, the metallic radii of Ni (124 pm) and Rh (134 pm) differ by $\sim 8$\%. Naturally, one would expect that, compared to pure Ni, there is more disorder in Ni$_{1-x}$Rh$_x$ alloy, making it a good candidate to test for the existence of a disorder-driven FM QCP. Polycrystalline Ni$_{1-x}$Rh$_x$ samples with $0.3\le x\le 0.42$ were prepared by arc-melting the constituents Ni and Rh and annealed at \ang{1000}~C. Magnetization measurements were carried out using a Quantum Design (QD) magnetic property measurement system. Zero-field muon spin relaxation measurements were performed at the M20 surface muon channel at TRIUMF. Specific heat was measured using a QD Dynacool physical property measurement system equipped with a dilution refrigerator. Thermal expansion was measured with a homemade capacitance dilatometer. More details about the sample characterizations and experimental methods are
provided in the Supplemental Material \cite{SM}.
 
 Figure~\ref{Fig1}(a) shows the $\mu_0 H$ = 0.01~T magnetic susceptibility $\Delta M(T)/H$ of Ni$_{1-x}$Rh$_x$, after a temperature-independent contribution $M_0$ was subtracted from the measured $M(T)$ ($\Delta M = M - M_0$). $\Delta M/H$ sharply increases as $T$ is lowered through $T_{\rm C}$ for $x = 0.32-0.36$ where $T_{\rm C}$ is determined both through a linear fit, as shown in Fig.~\ref{Fig1}(a), and the Arrott-Noakes analysis as discussed below. For $x_{crit} = 0.375$ (where $T_{\rm C} \rightarrow$ 0), $\Delta M/H$ shows only a small increase down to the lowest measured temperature of 2~K, consistent with the complete suppression of FM order. Isothermal magnetization measurements at $T = 2$~K confirm that Ni$_{1-x}$Rh$_x$ is a soft ferromagnet without a measurable hysteresis (Fig~\ref{Fig1}(b)). For the $x=0.32$ sample, which orders near 100~K, the inverse magnetic susceptibility $H/\Delta M$ exhibits Curie-Weiss-\textit{like} behavior between 150 and 300~K, from which we derive a paramagnetic (PM) effective moment $\mu_{\rm PM} = 1.97 \mu_{\rm B}$/f.u. (see SM). For the same sample, $\Delta M$ is small at 7~T ($\sim 0.22$ $\mu_{\rm B}$/f.u.), and the Rhodes-Wohlfarth ratio, $\mu_{\rm PM}/\mu_{sat}$ = 9, much larger than unity, is indicative of itinerant moment behavior in Ni$_{1-x}$Rh$_x$ \cite{Santiago2017}. An earlier study indicated spin glass behavior in Ni$_{1-x}$Rh$_x$ \cite{Carnegie1984}. However, our AC magnetic susceptibility measurements, presented in the SM, show no evidence for spin glass behavior near $T_{\rm C}$. Such a discrepancy may be due to different purity of starting materials or sample homogeneity. 

\begin{figure}[tbp]
		\centering
		\includegraphics[width=\columnwidth]{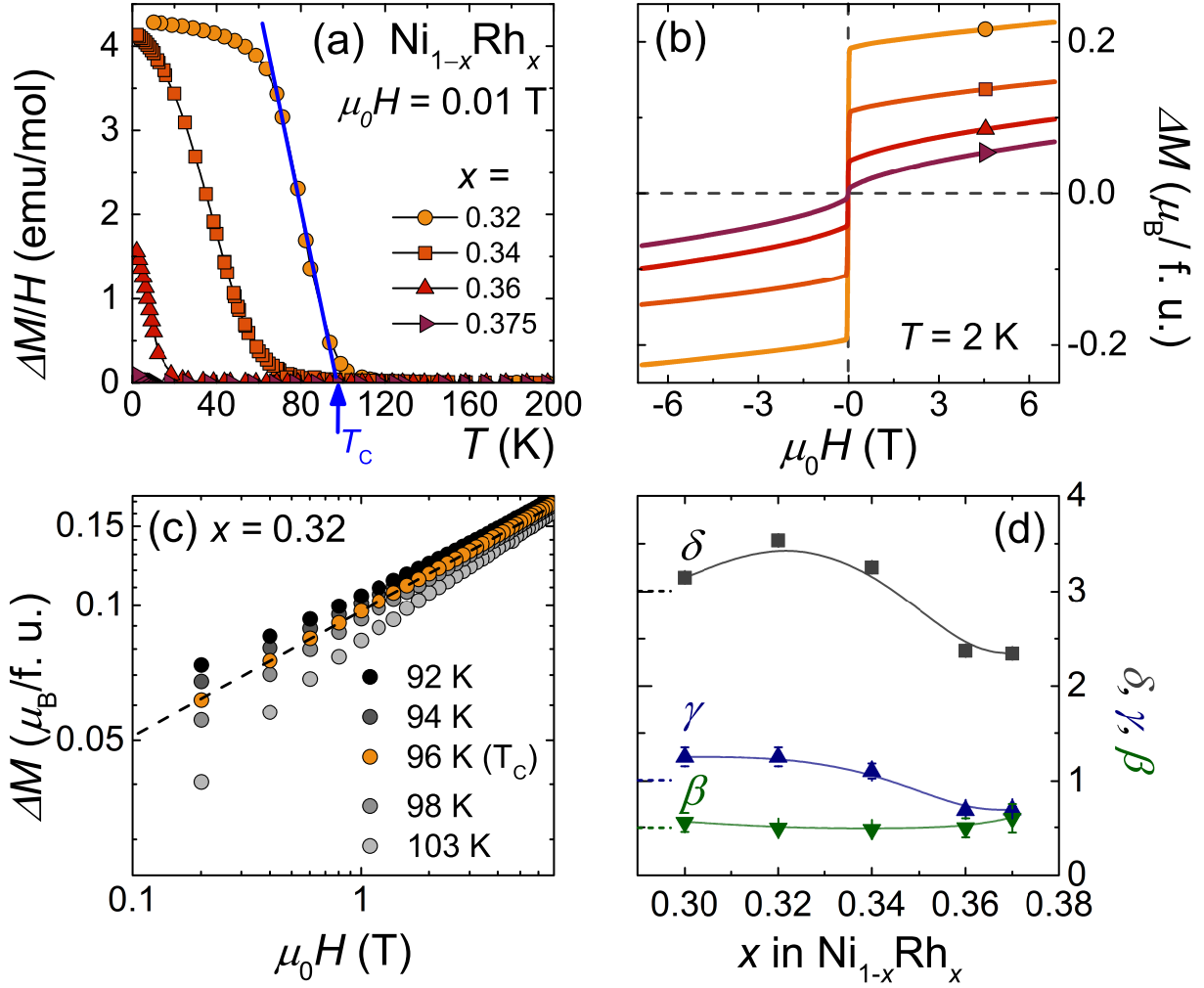}
		\caption{(a) Magnetic susceptibility $\Delta M/H = (M - M_0)/H$ for $\mu_0 H = 0.01$ T and (b) isothermal magnetization $\Delta M$ at $T = 2$~K of Ni$_{1-x}$Rh$_{x}$. Solid line in (a) shows how $T_{\rm C}$ was determined. (c) Log-log magnetization isotherms for $x = 0.32$, with the dashed line showing $T_{\rm C}$. (d) Critical exponents $\beta$, $\gamma$, and $\delta$ determined from the Arrott-Noakes scaling plots as a function of $x$. Solid lines are guides to the eye. Mean-field values $\beta = 0.5$, $\gamma = 1$, and $\delta = 3$ are indicated by horizontal dashed lines.}
		\label{Fig1}
	\end{figure}

For ferromagnets, the equation of state at $T_{\rm C}$ is given by $\Delta M \sim H^{1/\delta}$ \cite{Arrott1967}. From linear fits of log($\Delta M$) \textit{vs.} log ($\mu_{0}H$), as shown by the dashed line in Fig.~\ref{Fig1}(c), we determine that $T_{\rm C} = 96$~K and $\delta \sim 3.5$ for the $x=0.32$ sample. We applied the same analysis for all samples with $x = 0.30 - 0.37$. The critical exponents $\beta$ and $\gamma$ were determined by applying Arrott-Noakes scaling to the isotherms measured in the vicinity of $T_{\rm C}$ (see SM for details) \cite{Arrott1967}. The composition dependence of all three exponents, $\delta$, $\beta$, and $\gamma$, is summarized in Fig.~\ref{Fig1}(d). The Widom relation $\gamma/\beta = \delta -1$ is obeyed over the entire range of Rh concentrations investigated here, a self-consistent check of the scaling analysis. At $x = 0.30$, which is well below $x_{crit}$, the exponents $\beta = 0.5$, $\gamma = 1.3$, and $\delta = 3.1$ are close to the expected mean-field values. With increasing $x$, the exponents deviate from the mean-field values and approach $\beta = 0.6$, $\gamma = 0.7$, and $\delta = 2.3$ at $x = 0.37$, just below $x_{crit}$. A similar evolution of the critical exponents with chemical substitution was observed in Sr$_{1-x}$Ca$_x$RuO$_3$, where it was proposed that disorder resulted in enhanced quantum fluctuations near $x_{crit}$ \cite{Fuchs2014}.

	\begin{figure}
		\centering
		\includegraphics[width=\columnwidth]{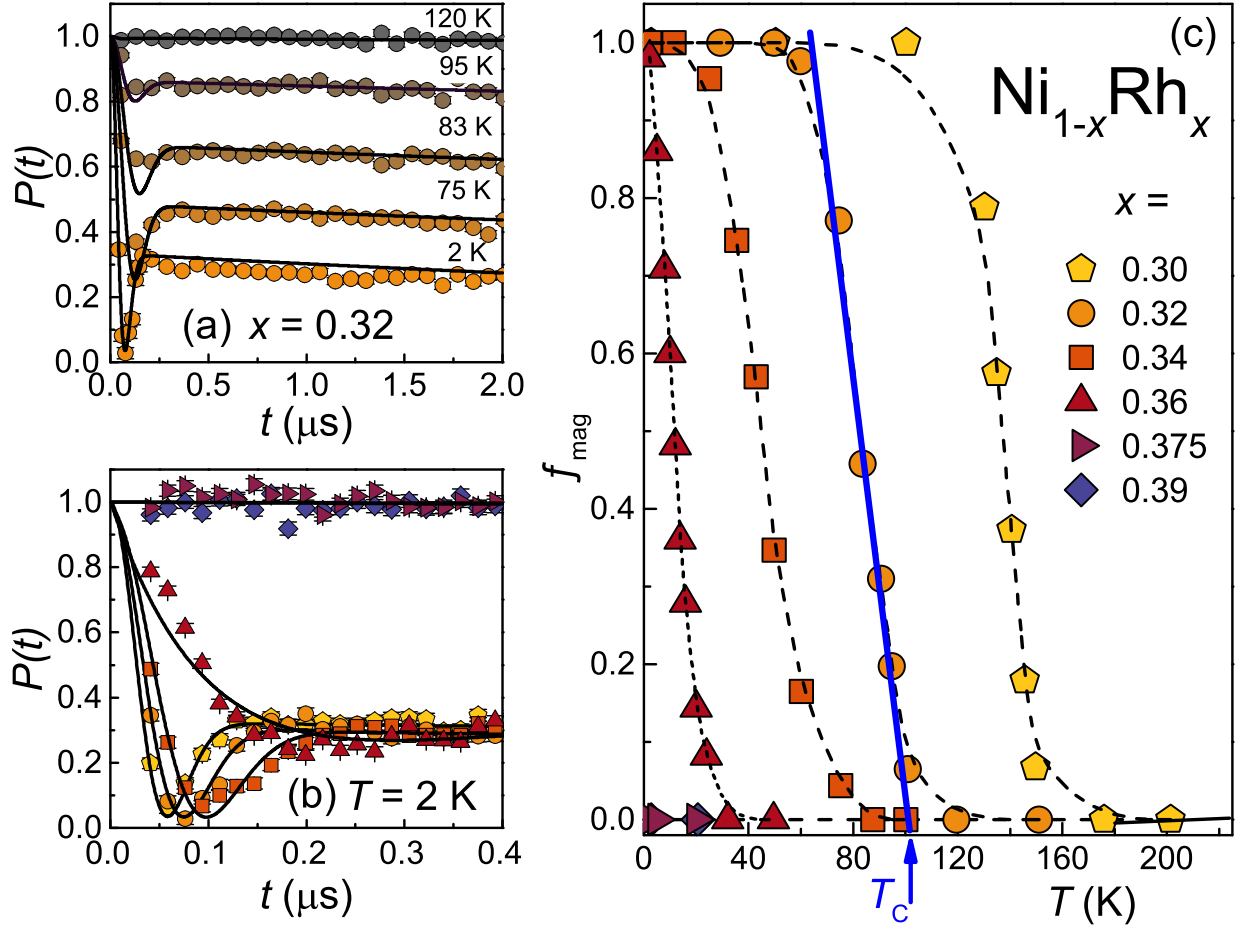}
		\caption{(a) Temperature evolution of the normalized muon decay asymmetry $P(t)$ for Ni$_{1-x}$Rh$_x$ for $x$ = 0.32. The solid lines are fits to Eqn.~\ref{eq:1}. (b) $P(t)$ for all measured samples $x = 0.30 - 0.39$, at $T = 2$~K. (c) The magnetic volume fraction $f_{mag}$ as a function of temperature. Solid line shows how $T_{\rm C}$ was determined.}
		\label{Fig2}
	\end{figure} 

Zero field $\mu$SR measurements were performed on six samples of Ni$_{1-x}$Rh$_x$ with $x = 0.30 - 0.39$, in order to determine whether the magnetic order takes place via a first- or second-order process. Hallmarks of a first-order transition are phase separation or an abrupt change of ground state \cite{Uemura2006,frandsen2016volume}. Conversely, in the case of a second-order transition, the size of the ordered moment is expected to continuously decrease without phase separation. $\mu$SR allows an independent measure of both the local order parameter and the magnetic volume fraction, $f_{mag}$, and can thus unambiguously distinguish between these scenarios. Representative muon decay asymmetry spectra, $P(t)$, are plotted in Fig.~\ref{Fig2}(a) for $x = 0.32$ at various temperatures below and above $T_{\rm C} = 96$~K. Above $T_{\rm C}$, $P(t)$ is essentially non-relaxing, as expected in a PM state. The onset of magnetic order is signaled by a fraction of the asymmetry undergoing rapid relaxation at early times. The compositional dependence of $P(t)$ at $T = 2$~K is presented in Fig.~\ref{Fig2}(b). This comparison reveals that the samples with the highest Rh concentrations, $x = 0.375$ and 0.39 ($\geq x_{crit}$, blue and purple symbols), exhibit only weak relaxation down to the lowest measured temperatures, thus confirming the absence of magnetic order for these compositions. The samples with $x < x_{crit}$ exhibit sharp relaxation associated with magnetic order. The $P(t)$ data for all compositions and temperatures is well-described by the dynamic Kubo-Toyabe function \cite{Kubo1967}:
\begin{equation}
P(t) = (1-f_{mag}) \cdot e^{-\lambda t} + f_{mag}\cdot G_{\text{DKT}}(t,\sigma,\nu)
\label{eq:1}
\end{equation}
where $\lambda$ and $\sigma$ are the relaxation rates for the non-magnetic and magnetic fractions of the sample, respectively, and $\nu$ is the hopping rate. The temperature dependence of $f_{mag}$ is presented in Fig.~\ref{Fig2}(c), revealing no evidence for phase separation; $f_{mag}$ remains 100\% up to Rh concentrations of $x = 0.36$ and drops to 0\% at $x_{crit} = 0.375$. With increasing Rh concentration, the Kubo-Toyabe minimum moves to increasing times as can be seen in Fig.~\ref{Fig2}(b), consistent with a decreasing ordered moment. This suggests that the suppression of magnetic order in Ni$_{1-x}$Rh$_x$ occurs via a continuous second-order process. 

\begin{figure}
		\centering
		\includegraphics[width=\columnwidth]{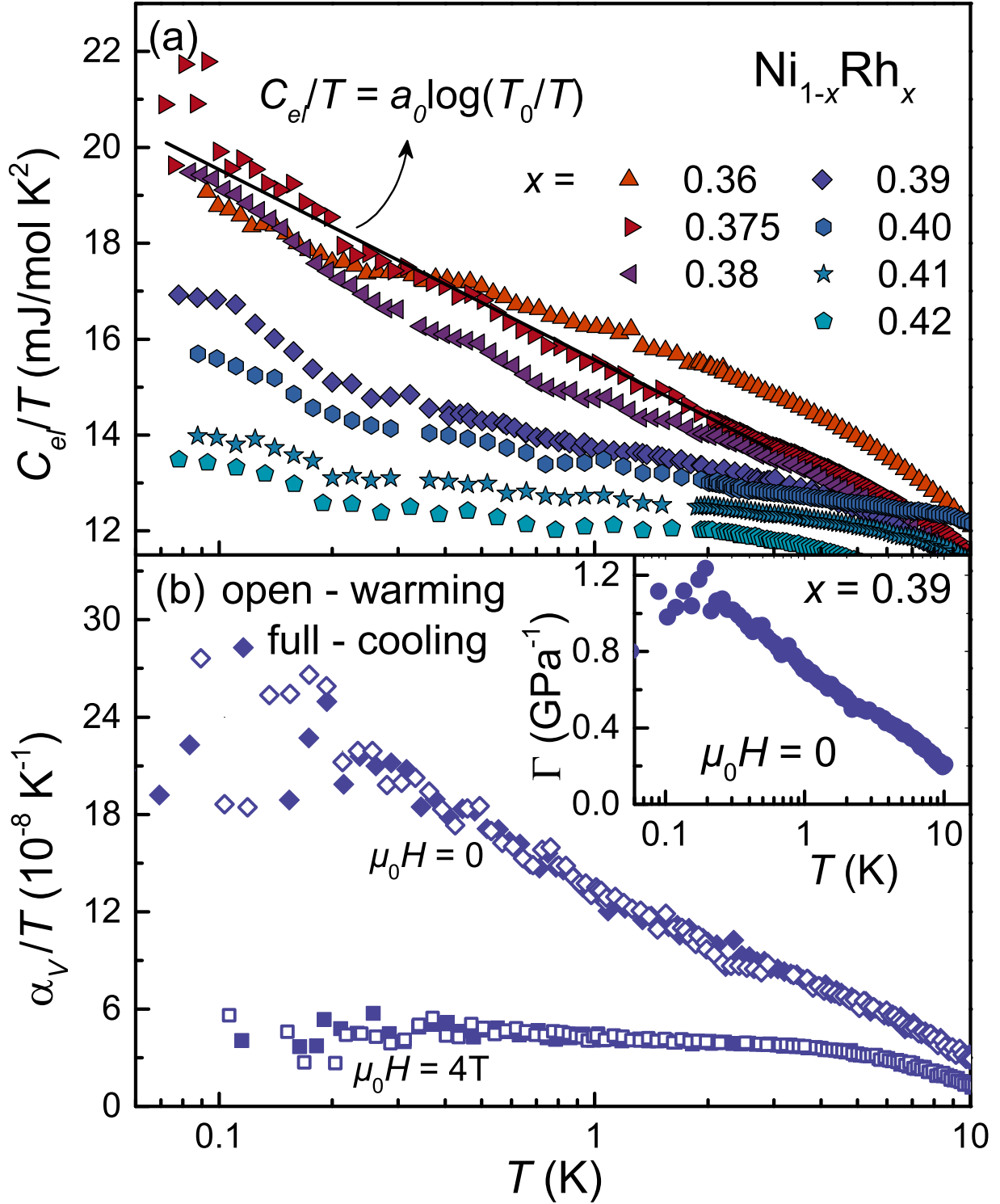}
		\caption{(a) Temperature dependence of the electronic specific heat $C_{el}/T$ for Ni$_{1-x}$Rh$_x$ with $x = 0.36-0.42$.  The solid line represents a fit to $C_{el}/T = a_0\log{(T_{0}/T)}$ at $x_{crit} = 0.375$. (b) The volume thermal expansion coefficient $\alpha_V/T$ at $\mu_0H=0$ (diamonds) and 4~T (squares) for Ni$_{1-x}$Rh$_x$ with $x = 0.39$. The inset shows the Gr\"uneisen ratio $\Gamma$ vs. $T$ at $\mu_0H=0$.}
		\label{Fig3}
	\end{figure}

\begin{figure}
		\centering
		\includegraphics[width=\columnwidth]{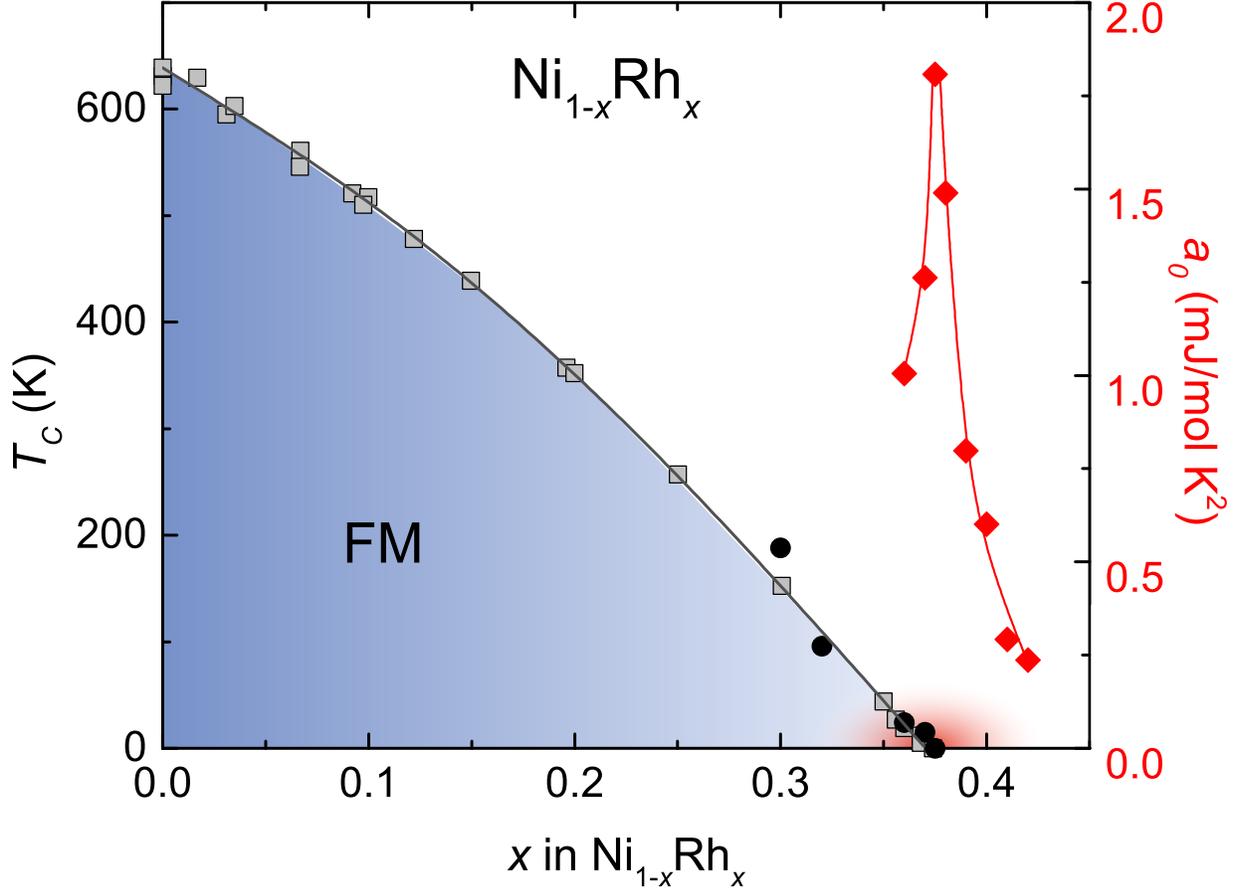}
		\caption{$T_{\rm C} - x$ phase diagram of Ni$_{1-x}$Rh$_x$. The blue region corresponds to long-range FM order. The red area marks the NFL behavior around the QCP. Black circles: $T_{\rm C}$ and red diamonds: the coefficient $a_0$ from the specific heat data (from current study). Gray squares: from Refs.~[\onlinecite{Muellner1975,Fujiwara1976,Vetter1981,Carnegie1984,Wijn1991}].}
		\label{Fig5}
	\end{figure}

Next we show evidence for divergent thermodynamic parameters in Ni$_{1-x}$Rh$_x$. Figure~\ref{Fig3}(a) shows the electronic specific heat  $C_{el}/T$ around $x_{crit}=0.375$, where the phonon contribution has been subtracted from the measured specific heat. For concentrations that are both far above and far below $x_{crit}$ ($x \leq 0.15$ and $x \geq 0.6$), $C_{el}/T$ is nearly temperature-independent at low temperatures, as expected for a Fermi liquid (FL) \cite{SM}. Close to $x_{crit}$, $C_{el}/T$ diverges logarithmically on cooling. The fastest divergence occurs at $x_{crit} = 0.375$, where $C_{el}/T = a_0\log{(T_{0}/T)}$ between 0.1 and 3~K (solid line in Fig.~\ref{Fig3}(a)), such that $a_0$ is maximum at the QCP (red diamonds in Fig.~\ref{Fig5}). This logarithmic divergence was previously reported in Ni$_{0.62}$Rh$_{0.38}$ \cite{Triplett1971} and has also been observed in other QCP systems \cite{Brando2008,Steppke2013,Jia2011,Svanidze2015}. For $x > x_{crit}$, $C_{el}/T$ levels off at the lowest temperatures, consistent with non-Fermi-liquid (NFL) to FL crossover. This is similar to other FM and antiferromagnetic quantum critical systems  \cite{Stewart2001,Stewart2006,Loehneysen2007,Manuel2016}.

QCPs are characterized by an accumulation of magnetic entropy $S_{mag}$ as a function of the control parameter at low, but finite temperatures. 
In Ni$_{1-x}$Rh$_x$, this is underscored by the dependence of the specific heat parameter $a_0$ on $x$ (red diamonds in Fig.~\ref{Fig5}), given that $S_{mag}$ is commensurate to $a_0$, which, in turn, is maximum at the QCP. At the same time, $S_{mag}$ is related to the volume thermal expansion $\alpha_V$ through the Maxwell relation $\alpha_V = -V^{-1} \partial S_{mag}/\partial p$ (where $p$ is pressure), and the divergence of $\alpha_V/T$ has been taken as proof of the QCP in heavy fermion systems, such as CeCu$_{6-x}$Au$_{x}$ \cite{Grube2017}, CeNi$_2$Ge$_2$, and YbRh$_2$(Si$_{0.95}$Ge$_{0.05}$)$_2$ \cite{Kuechler2003}. Our data shows that at $x = 0.39$ (just above the QCP), zero-field $\alpha_{V}/T$ diverges logarithmically between 10 and 0.1~K (diamonds in Fig.~\ref{Fig3}(b)). This is indicative of NFL behavior in proximity to the QCP \cite{Zhu2003}. The data show no hysteresis between heating (open) and cooling (full) measurements, ruling out any history-dependent spin glass effects. The length measurements on Ni$_{1-x}$Rh$_x$ with $x = 0.39$ reached the resolution limit of the dilatometer of $\Delta L \geq 10^{-3}$\,\AA\ at the lowest measured temperatures, resulting in an enhanced scattering below $\sim 0.2$ K. The application of a magnetic field of 4~T reduces $\alpha_V/T$ to a nearly constant value below 4~K, indicating a recovery of the FL behavior (squares in Fig. \ref{Fig3}(b)). This recovery of FL behavior is consistent with what has been observed in field-dependent specific heat measurements for Ni$_{0.62}$Rh$_{0.38}$ \cite{Triplett1971}. 

An additional probe for a QCP is the Gr\"uneisen ratio $\Gamma = \alpha_V/C_{el} \sim 1/E^* \cdot \partial E^*/\partial p$. $\Gamma$ reveals the hydrostatic pressure dependence of the dominating, characteristic energy scale $E^*$ (e.g., the energy related to the conduction band splitting at the Fermi energy, which is proportional to the spontaneous magnetization \cite{Mohn2006}). At a QCP, $E^*$ vanishes, and $\Gamma$ is expected to diverge with decreasing $T$ \cite{Zhu2003}. In the low temperature range for the $\alpha_V$ measurements, the phonon contribution is negligible. The calculated $\Gamma$ is depicted in the inset of Fig. \ref{Fig3}(b), showing logarithmic divergence over two decades in temperature from $T = 10$~K to 0.1~K. The fact that $\Gamma \sim -$ log$T$ suggests either that the quantum critical behavior in Ni$_{1-x}$Rh$_x$ extends to a finite pressure interval (rather than a point) \cite{Zhu2003}, or that the system lies within a disordered quantum Griffiths phase \cite{Vojta2009}.

We summarize the $T_{\rm C} - x$ phase diagram of Ni$_{1-x}$Rh$_x$ in Fig.~\ref{Fig5}. Magnetization $M(T,H)$ and $\mu$SR measurements reveal the suppression of $T_C$ with increasing Rh concentration up to $x_{crit} = 0.375$ (black symbols). The magnetically-ordered volume fraction remains 100\% up to $x_{crit}$, while the magnitude of the ordered moment per formula unit continuously decreases, as expected for a second order transition \cite{Goko2017}. In addition, the FM QCP is also revealed by the divergence of $C_{el}/T$, $\alpha_V/T$, and $\Gamma$ in the low temperature limit, associated with NFL behavior that extends up to $\sim 10$~K. 

Finally, we compare our results with other Ni$_{1-y}M_{y}$ ($M$ = Al, Si, V, Cr, Mn, Cu, Zn, Pd, and Sb) alloys. Nonmagnetic $M$ metals dilute the Ni magnetic moment and therefore suppress the FM order. Magnetic susceptibility measurements on these alloys are sensitive to sample preparation \cite{Gregory1975,Carnegie1984}. In the absence of a spin glass state or short range order, the enhancement of $C_{el}/T$ has been observed for all $M$ where $T_{\rm C} \rightarrow 0$ \cite{Gupta1964,Gregory1975,Nicklas1999}. This commonality can be understood in terms of enhanced spin fluctuations and does not necessarily indicate quantum critical fluctuations. A noteworthy member of this family is Ni$_{1-y}$V$_y$ where V substitution results in quantum Griffiths effect that competes with critical behavior without reaching a QCP \cite{Ubaid-Kassis2010,Vojta2010}. By contrast, Ni$_{1-x}$Rh$_x$ is the first member of the Ni$_{1-y}M_{y}$ family where divergent $\alpha_{V}/T$ and $C_{el}/T$ result in divergent $\Gamma$ \cite{Zhu2003}, demonstrating the presence of a FM QCP. In fact, for most ferromagnets, when a dilution occurs in the magnetic sublattice, short-range order or spin glass behavior is observed \cite{Manuel2016}. The only exception is the 5$f$-electron system Th$_{1-x}$U$_{x}$Cu$_2$Si$_2$ that the FM transition remains continuous at the critical concentration, where NFL behavior is observed \cite{Lenkewitz1997}. 

One plausible scenario to account for the FM QCP in Ni$_{1-x}$Rh$_x$ is the aforementioned BKV theory \cite{Belitz1999PRL,Sang2014,Kirkpatrick2015}. The current study utilized polycrystalline samples and the residual resistivity ratio (not shown), which is often taken as a gauge of the amount of disorder, is small and comparable among the whole series of Ni$_{1-x}$Rh$_x$. To test if the FM quantum criticality in Ni$_{1-x}$Rh$_x$ fulfills the universality class in the strong disorder regime of the BKV theory, the growth of single crystals is imperative and is the subject of an ongoing study. Ni$_{1-x}$Rh$_x$ shows the first occurrence of a FM QCP with dilution of the $d$-electron magnetic sublattice. This is in contrast with chemical substitution on the non-magnetic sublattice in other FM QCP systems \cite{Jia2011,Steppke2013,Huang2016,Goko2017,Sales2017,Lai2018}. In particular, due to its chemical simplicity, Ni$_{1-x}$Rh$_x$ is an ideal platform for furture studies and our work establishes a new approach to explore FM quantum criticality.

\begin{acknowledgments}
We acknowledge V. Taufour, D.-N. Cho, and D. Belitz for fruitful discussions. The work at Rice University was funded by the NSF DMR 1903741. We thank G. Costin for his assistance with EPMA measurements. The use of the EPMA facility at the Department of Earth, Environmental and Planetary Sciences, Rice University, is kindly acknowledged. We are grateful to Bassam Hitti and Gerald Morris for their assistance with the muon spin relaxation measurements. Research at McMaster University is supported by the Natural Sciences and Engineering Research Council of Canada. T.B. and T.S. are supported by award DE-SC0008832 from the Materials Sciences and Engineering Division in the U.S. Department of Energy's Office of Basic Energy Sciences and the National High Magnetic Field Laboratory through the NSF Cooperative Agreement No. DMR-1157490 and the State of Florida.
\end{acknowledgments}

%

\end{document}